\documentclass[conference,10pt]{IEEEtran}
\IEEEoverridecommandlockouts
\usepackage{amsmath,amsfonts,amsthm,amssymb,bbm}
\usepackage{cite}
\usepackage{balance}
\usepackage{xcolor}
\definecolor{asparagus}{rgb}{0.53, 0.66, 0.42}
\definecolor{armygreen}{rgb}{0.29, 0.33, 0.13}
\definecolor{airforceblue}{rgb}{0.36, 0.54, 0.66}
\definecolor{brown(web)}{rgb}{0.65, 0.16, 0.16}
\definecolor{burgundy}{rgb}{0.5, 0.0, 0.13}
\definecolor{darkcerulean}{rgb}{0.03, 0.27, 0.49}
\definecolor{indigo(dye)}{rgb}{0.0, 0.25, 0.42}
\usepackage{mathrsfs}
\usepackage{xcolor}
\usepackage{graphicx}

\theoremstyle{plain} 
\newtheorem{theorem}{Theorem}

\newtheorem{definition}{Definition}
\newtheorem{lemma}{Lemma}

\theoremstyle{definition} \newtheorem{remark}{Remark}
\theoremstyle{definition} 
 
\allowdisplaybreaks
\usepackage{dblfloatfix}
\newcounter{MYtempeqncnt}

\def \extended {1} % defines a variable 'extended' which can take a value either 0 or 1. Toggling to 0 gives 6 page version and toggling to 1 gives extended version on compilation.
\title{An Alphabet of Leakage Measures}
\author{Atefeh Gilani, Gowtham R. Kurri, Oliver Kosut, Lalitha Sankar
\thanks{The authors are with the School of Electrical, Computer and Energy Engineering at Arizona State University. Email: {\tt agilani2@asu.edu, gkurri@asu.edu, okosut@asu.edu, lalithasankar@asu.edu}}
\thanks{This work is supported in part by NSF grants CIF-1901243, CIF-1815361, CIF-2007688, CIF-2134256, and SaTC-2031799.}}

\begin{document}
\maketitle
\begin{abstract}
We introduce a family of information leakage measures called \emph{maximal $\alpha,\beta$-leakage}, parameterized by real numbers $\alpha$ and $\beta$. The measure is formalized via an operational definition involving an adversary guessing an unknown function of the data given the released data. We obtain a simple, computable expression for the measure and show that it satisfies several basic properties such as monotonicity in $\beta$ for a fixed $\alpha$, non-negativity, data processing inequalities, and additivity over independent releases. Finally, we highlight the relevance of this family by showing that it bridges several known leakage measures, including maximal $\alpha$-leakage $(\beta=1)$, maximal leakage $(\alpha=\infty,\beta=1)$, local differential privacy $(\alpha=\infty,\beta=\infty)$, and local R\'{enyi} differential privacy $(\alpha=\beta)$. 
\end{abstract}

\section{Introduction}\label{intro}
% A measure of information leakage aims to quantify the amount of information leaked/revealed about some \emph{sensitive data} to an adversary who observes a correlated \emph{released data}. Various information leakage measures have been proposed in the literature. 

% Various measure of information leakage

How much information does an observation released to an adversary reveal/leak about correlated sensitive data? This fundamental question arises in many secrecy and privacy problems whenever data about users is stored online (e.g., social networks and cloud-based services) and a certain level of information leakage is unavoidable in exchange for certain services. %The release of an observation to an adversary could be either an inadvertent result of a design via a side channel or intentional, e.g., in the context of querying databases. 
In an effort to quantify this leakage precisely, a variety of privacy measures have been proposed~\cite{RassouliDeniz,asoodeh2015maximal,AsoodehDL2017,mironov2017renyi,IssaWK2020,LiaoKS20,Dwork2006,kasiviswanathan2011can,duchi2013local}.

 For any leakage measure, one of the key challenges is to associate an operational interpretation in terms of its definition. Only a few leakage measures possess such an operational meaning. For example, the works in \cite{AsoodehDL2017,IssaWK2020,LiaoKS20}, which pertain to the release of observation due to a side channel, measure privacy in terms of an adversary's gain in \emph{guessing} the sensitive data after observing the released data. %Asoodeh \emph{et al.}~\cite{AsoodehDL2017} use the probability of correctly guessing to measure privacy. 
 Issa \emph{et al.}~\cite{IssaWK2020} introduce maximal leakage (MaxL), which quantifies the maximal logarithmic gain in the probability of correctly guessing any arbitrary function of the original data from the released data. Liao \emph{et al.}~\cite{LiaoKS20} later generalized maximal leakage to a family of leakages, maximal $\alpha$-leakage (Max-$\alpha$L), that allows tuning the measure to specific applications. In particular, similar to MaxL, Max-$\alpha$L quantifies the maximal logarithmic gain in a monotonically increasing power function (dependent on $\alpha$) applied to the probability of correctly guessing. 
 
 Among leakage measures motivated by worst-case adversaries, differential privacy (DP)~\cite{Dwork2006} has emerged as the gold standard. A differentially private algorithm guarantees that its outputs restrict the adversary from distinguishing between neighboring data entries. When privacy guarantees have to be provided in a distributed setting, local DP (LDP)~\cite{kasiviswanathan2011can,duchi2013local} provides such strong guarantees for every pair of (users) data entries. In the context of composing DP outputs sequentially, R\'{e}nyi differential privacy (RDP)~\cite{mironov2017renyi} has emerged as a meaningful variant to compute overall DP guarantees. Specifically, RDP relaxes DP based on the R\'{e}nyi divergence. 
 
 %and its variants, local differential privacy (LDP)~\cite{kasiviswanathan2011can,duchi2013local}, and R\'{e}nyi differential privacy (RDP)~\cite{mironov2017renyi},
 %In particular, DP guarantees indistinguishability of query outputs over neighboring datasets (thus requiring all data accessible centrally), whereas LDP is a distributed model with stronger privacy guarantees where each individual user randomizes their query output. %An operational interpretation of DP in the framework of hypothesis testing is given by Kairouz \emph{et al.}\cite{kairouz2015composition}, where they show that it determines the trade-off between probabilities of false alarm and missed detection.
 
No single measure of privacy/information leakages suits all the scenarios in practice. In this paper, we undertake the study of unifying various measures of information leakage so that the leakage measure can be tailored to different settings depending on the context. Motivated by \cite{IssaWK2020,LiaoKS20}, in Section \ref{sec:Theorems-maximal alpha-beta leakage}, we introduce a leakage measure, \emph{maximal $\alpha,\beta$-leakage}, which is parameterized by two real numbers $\alpha$ and $\beta$. We obtain a simple computable expression for it and show that this family of measures encompasses a host of existing leakage measures: in particular, Max-$\alpha$L ($\beta=1$), MaxL ($\alpha=\infty,\beta=1$), LDP ($\alpha=\beta=\infty$), and local R\'{e}nyi differential privacy (LRDP) ($\alpha=\beta$) --- a notion of RDP defined analogous to LDP (see Fig. \ref{fig:alpha_beta_leakage_and_other_measures}). An important consequence of our result is an operational interpretation of LDP and LRDP (Section \ref{sec:relationship}). We note that this subsumes an operational meaning of LDP given by Issa \emph{et al.}~\cite{IssaWK2020} via maximal realizable leakage, a leakage measure definition concerned with worst-case analysis, akin to LDP. Interestingly, maximal $\alpha,\beta$-leakage is defined in terms of average-case analysis (in the spirit of MaxL and Max-$\alpha$L), and yet, it recovers the worst-case LDP by exploiting the interplay between the parameters $\alpha$ and $\beta$. We also show that this general leakage measure satisfies all the axiomatic properties of a measure of information leakage, including non-negativity, equality to zero if and only if the original data and the released data are independent of each other, and data-processing inequalities. This can be viewed as another proof that LDP satisfies both the post-processing and linkage inequalities unlike DP which does not satisfy the linkage inequality~\cite{Basciftietal}. 
\if\extended 0 
Due to space constraints, we have omitted some proofs; detailed proofs for all results are in \cite{extendedITW2022-Gilani}.
\fi
\section{Preliminaries}
We begin by reviewing the definitions of some existing information leakage measures, in particular, maximal $\alpha$-leakage (which subsumes maximal leakage) and local (R\'{enyi}) DP.

\begin{definition}[Maximal $\alpha$-leakage~\cite{LiaoKS20}]
Let $P_{XY}$ be a joint distribution, where $X$ and $Y$ represent the original data and the released data, respectively. The maximal $\alpha$-leakage from $X$ to $Y$, for $\alpha \in (1,\infty)$, is defined as 
\begin{align}\label{eqn:max-alpha-leakage-def}
    %  \mathcal{L}_\alpha^{\emph{max}}(X \to Y)\!= \!\sup_{U-X-Y}\frac{\alpha}{\alpha-1} \!\log \frac{\displaystyle \max_{P_{\hat{U}|Y}} \ \mathbb{E}\left[ P_{\hat{U}|Y}(U|Y)^{\frac{\alpha-1}{\alpha}}\right]}{\displaystyle\max_{P_{\hat{U}}}\mathbb{E}\left[P_{\hat{U}}(U)^{\frac{\alpha-1}{\alpha}}\right]}
    \mathcal{L}&_\alpha^{\emph{max}}(X \to Y)=\frac{\alpha}{\alpha-1}\nonumber\\
     &\sup_{U-X-Y}\log\frac{\max\limits_{P_{\hat{U}|Y}}\sum_{u,y}P_{UY}(u,y)P_{\hat{U}|Y}(u|y)^{\frac{\alpha-1}{\alpha}}}{\max\limits_{P_{\hat{U}}}\sum_uP_U(u)P_{\hat{U}}(u)^{\frac{\alpha-1}{\alpha}}},
\end{align}
%where $U$ represents any randomized function of $X$ that takes values in a arbitrary finite alphabet and $\hat{U}$ is an estimator of $U$ with the same support as $U$.
where $U$ represents any randomized function of $X$ that the adversary is interested in guessing and takes values in an arbitrary finite alphabet. Moreover,  $\hat{U}$ is an estimator of $U$ with the same support as $U$.
\end{definition}
Maximal $\alpha$-leakage is a generalization of another measure of information leakage, the maximal leakage~\cite{IssaWK2020}. In particular, the latter recovers the former when $\alpha=\infty$. Liao~\emph{et~al.}~\cite{LiaoKS20} showed that 
 \begin{align}\label{eqn:maximalalpha}
     \mathcal{L}_\alpha^{\text{max}}(X\rightarrow Y)=\sup_{P_{\tilde{X}}}I_\alpha^{\text{S}}(\tilde{X};Y),
 \end{align}
 where the supremum is over all the probability distributions $P_{\tilde{X}}$ on the support of $P_X$ and $I_\alpha^\text{S}(\cdot;\cdot)$ is the Sibson mutual information of order $\alpha$~\cite{sibson1969information}.

\begin{definition}[Local differential privacy~\cite{kasiviswanathan2011can,duchi2013local}]
Given a conditional distribution $P_{Y|X}$, the local differential privacy (LDP) is defined as
\begin{align}
    \mathcal{L}^{\emph{LDP}}(X\rightarrow Y)=\max_{\substack{y\in\mathcal{Y},\\ x,x^\prime\in\mathcal{X}}}\log{\frac{P_{Y|X}(y|x)}{P_{Y|X}(y|x^\prime)}}.
\end{align}
\end{definition}
We may define local R\'{e}nyi differential privacy as a generalization of local differential privacy based on the R\'{e}nyi divergence~\cite{renyi1961measures}.
\begin{definition}[Local R\'{e}nyi differential privacy]
Given a conditional distribution $P_{Y|X}$, the local R\'{e}nyi differential privacy (LRDP) is defined as
\begin{align}
    \mathcal{L}&^{\emph{LRDP}}(X\rightarrow Y)\nonumber\\
    &=\max_{x,x^\prime\in\mathcal{X}}\frac{1}{\alpha-1}
 \log \sum_y P_{Y|X}(y|x')^{1-\alpha} P_{Y|X}(y|x)^\alpha. 
\end{align}
\end{definition}
It can be verified using L'H\^{o}pital's rule that LRDP simplifies to LDP as $\alpha\rightarrow \infty$.
\section{Maximal $\alpha,\beta$-Leakage}\label{sec:Theorems-maximal alpha-beta leakage}
Motivated by the definitions of maximal leakage and maximal $\alpha$-leakage, we introduce maximal $\alpha,\beta$-leakage as follows.
\begin{definition}[Maximal $\alpha,\beta$-leakage] Given a conditional distribution $P_{Y|X}$, the maximal $\alpha,\beta$-leakage from $X$ to $Y$ for $\alpha \in (1,\infty)$ and $\beta \in [1,\infty)$ is defined as

\begin{align}\label{eqn:alpha,beta-leakage-original-def}
   &\nonumber\mathcal{L}_{\alpha,\beta}(X\to Y):=\sup_{P_{X}}\ \sup_{U\to X\to Y}\  \frac{\alpha}{\alpha-1}
\\ &
 \log \frac{\displaystyle \max_{P_{\hat{U}|Y}} \left[\sum_y P_Y(y) \left(\sum_u P_{U|Y}(u|y) P_{\hat{U}|Y}(u|y)^{\frac{\alpha-1}{\alpha}}\right)^{\beta}\right]^{1/\beta}}{\displaystyle \max_{P_{\hat{U}}} \sum_u P_U(u)P_{\hat{U}}(u)^{\frac{\alpha-1}{\alpha}}}.
\end{align}
where $\hat{U}$ represents an estimator taking values from the same arbitrary finite alphabet as $U$. It is defined by continuous extension for $\alpha=\infty$ or $\beta=\infty$.
% , and is given by.
% % \begin{align}\label{eqn:alphabeta-inf-inf}
% % &\mathcal{L}_{\infty,\infty}(X\rightarrow Y)\nonumber\\
% % &=\sup_{P_X}\sup_{U-X-Y}\log\frac{\max\limits_{y\in\emph{supp}(Y)}\max\limits_{P_{\hat{U}|Y}}\sum\limits_{u}P_{U|Y}(u|y)P_{\hat{U}|Y}(u|y)}{\max\limits_{P_{\hat{U}}}\sum\limits_uP_{U}(u)P_{\hat{U}}(u)}    
% % \end{align}
% \begin{align}\label{eqn:alphabeta-inf-inf}
% \mathcal{L}_{\infty,\infty}(X\rightarrow Y)=\lim_{\alpha,\beta\rightarrow \infty}\mathcal{L}_{\alpha,\beta}(X\to Y).
% \end{align}
\end{definition}
 We remark that the definition of maximal $\alpha,\beta$-leakage in \eqref{eqn:alpha,beta-leakage-original-def} nearly recovers the definition of maximal $\alpha$-leakage from \eqref{eqn:max-alpha-leakage-def} (and thus maximal leakage also) when $\beta=1$. The reason is that maximal $\alpha$-leakage depends on the distribution of $X$ only through its support, as shown in \eqref{eqn:maximalalpha}, so including the supremum over $P_X$ does not change the value. Moreover, for $\alpha=1$, maximal $\alpha,\beta$-leakage recovers Shannon channel capacity.
%rather than mutual information.}
%\noteGK{The one in red below is more easily parsable to me than the blue one.}
%\textcolor{red}{While the definition of Max-$\alpha$L does not include the supremum over $P_X$, as shown in \eqref{eqn:max-alpha-leakage-def}, Max-$\alpha$L has an implicit supremum over $P_X$ \noteGK{(in \eqref{eqn:maximalalpha})}, so including this supremum does not change the value.} \noteGK{How about adding the line here - ``Moreover, for $\alpha=1$, maximal $\alpha,\beta$-leakage recovers Shannon channel capacity."}
We have included the supremum in the definition of maximal $\alpha,\beta$-leakage in order to recover some of the worst-case measures such as LDP and LRDP as special cases. One can view the introduction of $\beta$ into the summation in the numerator in \eqref{eqn:alpha,beta-leakage-original-def} as allowing a continuous transition from a simple average over $y$ (at $\beta=1$) to a maximum over $y$ (at $\beta=\infty$). This maximum over $y$ is present in the definition of maximal \emph{realizable} leakage~\cite[Definition~8]{IssaWK2020}, which corresponds to $\alpha=\beta=\infty$, and has been shown to be equal to LDP. This allows us to view maximal leakage and maximal realizable leakage as two corner points of the inner optimization problem in \eqref{eqn:alpha,beta-leakage-original-def} for $\alpha=\infty$ and $\beta=1$ and $\beta=\infty$, respectively (see also Fig.~\ref{fig:alpha_beta_leakage_and_other_measures}). 
 %$\alpha=\infty,\beta=1$ and $\alpha=\infty,\beta=\infty$, respectively. 

The following theorem simplifies the expression of maximal $\alpha,\beta$-leakage in \eqref{eqn:alpha,beta-leakage-original-def}.
\begin{theorem}\label{thm:alpha-beta-leakage}
Maximal $\alpha,\beta$-leakage defined in \eqref{eqn:alpha,beta-leakage-original-def} simplifies to
\begin{align}\label{thm1}
&\nonumber\mathcal{L}_{\alpha,\beta}(X\to Y)
=\max_{x'} \  \sup_{P_{\Tilde{X}}}\frac{\alpha}{(\alpha-1)\beta} \\ &   \log 
\sum_y P_{Y|X}(y|x')^{1-\beta} \left(\sum_{x} P_{\Tilde{X}}(x) P_{Y|X}(y|x)^\alpha \right)^{\beta/\alpha},
\end{align}
where $P_{\Tilde{X}}$ is a probability distribution on the support of $P_{X}$.
\end{theorem}

A detailed proof for Theorem~\ref{thm:alpha-beta-leakage} is given in Section~\ref{proof:thm1}. 
% A simple corollary of this theorem is that $\mathcal{L}_{\infty,\infty}(X\rightarrow Y)$ in \eqref{eqn:alphabeta-inf-inf} also involves supremum of maximal realizable leakage, thereby leading to the equivalence of two ways of defining it (pertaining to the order of the limit and both the supremums).
For $\beta\le\alpha$, the quantity inside the log in \eqref{thm1} is concave in $P_{\tilde{X}}$; thus the supremum over $P_{\tilde{X}}$ can be efficiently solved using convex optimization techniques. As we will show in Section~\ref{sec:relationship}, for $\beta\ge\alpha$, the supremum over $P_{\tilde{X}}$ can be replaced by a maximum over $x\in\mathcal{X}$. Thus, in either case the quantity in \eqref{thm1} can be efficiently computed for finite alphabets.

\begin{figure*}[b]
% ensure that we have normalsize text
\normalsize
% Store the current equation number.
\setcounter{MYtempeqncnt}{\value{equation}}
% Set the equation number to one less than the one
% desired for the first equation here.
% The value here will have to changed if equations
% are added or removed prior to the place these
% equations are referenced in the main text.
\hrulefill
\setcounter{equation}{15}
\begin{equation}
\label{n_length_extension}
%\mathcal{L}_{\alpha,\beta}(X^n\to Y):=
\sup_{P_{X^n}} \ \max_i \ \sup_{U\to X^n\to Y} \frac{\alpha}{(\alpha-1)\beta}\; \log
\frac{ \displaystyle\sup_{P_{\hat{U}|X_{-i},Y}} \sum_{x_{-i},y} P_{X_{-i},Y}(x_{-i},y) \left[\sum_u P_{U|X_{-i},Y}(u|x_{-i},y)P_{\hat{U}|X_{-i},Y}(u|x_{-i},y)^{\frac{\alpha-1}{\alpha}}\right]^{\beta}}{\displaystyle\sup_{P_{\hat{U}|X_{-i}}}\sum_{x_{-i}} P_{X_{-i}}(x_{-i}) \left[\sum_u P_{U|X_{-i}}(u|x_{-i}) P_{\hat{U}|X_{-i}}(u|x_{-i})^{\frac{\alpha-1}{\alpha}}\right]^\beta}.
\end{equation}
% Restore the current equation number.
\setcounter{equation}{\value{MYtempeqncnt}}
% The IEEE uses as a separator
% The spacer can be tweaked to stop underfull vboxes.
\vspace*{4pt}
\end{figure*}

Like other leakage measures, maximal $\alpha,\beta$-leakage satisfies several basic properties such as non-negativity, data processing inequalities and additivity, as shown in the following theorem.
\begin{theorem}\label{thm2}
For $\alpha\in (1,\infty)$ and $\beta \in [1,\infty)$, maximal $\alpha,\beta$-leakage 
\begin{enumerate}
 \item is monotonically non-decreasing in $\beta$ for a fixed $\alpha$;
  \item satisfies data processing inequalities, i.e., for the Markov chain $X -Y - Z$:
    \begin{subequations}
            \begin{equation}\label{data-processing2}
        \mathcal{L}_{\alpha,\beta}(X\to Z) \leq \mathcal{L}_{\alpha,\beta}(X \to Y)
    \end{equation}
    \begin{equation}\label{data-processing1}
        \mathcal{L}_{\alpha,\beta}(X\to Z) \leq \mathcal{L}_{\alpha,\beta}(Y \to Z).
        \end{equation}
    \end{subequations}
     \item is non-negative, i.e.,
    \begin{align}
        \mathcal{L}_{\alpha,\beta}(X\to Y) \geq 0
    \end{align}
with equality if and only if $X$ and $Y$ are independent.
%\item satisfies additivity: i.e., if $(X_1,Y_1)$ and $(X_2,Y_2)$ are independent, then
\item satisfies additivity: i.e., if $(X_i,Y_i)$ for $i=1,2,\ldots,n$ are independent, then
% \begin{align}\label{eqn:additivity}
%   \mathcal{L}_{\alpha,\beta}(X_1,X_2 \to Y_1,Y_2)=\sum_{i\in\{1,2\}}\mathcal{L}_{\alpha,\beta}(X_i \to Y_i).  
\begin{align}\label{eqn:additivity}
  \mathcal{L}_{\alpha,\beta}(X_1,\ldots,X_n \to Y_1,\ldots,Y_n)=\sum_{i=1}^n\mathcal{L}_{\alpha,\beta}(X_i \to Y_i).  
\end{align}
\end{enumerate}
\end{theorem}
\if \extended 0
Proofs for most of the claims of Theorem~\ref{thm2} are in Section~\ref{proof:thm2}.
\fi
\if \extended 1
A detailed proof of Theorem~\ref{thm2} is in Section~\ref{proof:thm2}.
\fi
\begin{remark}\label{remark:reparameterization}
Although maximal $\alpha,\beta$-leakage is monotonic in only one of its orders, if we consider a reparameterization in which 
 $\tau \in [0,1]$ and $\beta=\frac{\alpha}{1-\tau(1-\alpha)}$, the new leakage measure is non-increasing in $\tau$ for a fixed $\alpha$, and non-decreasing in $\alpha$ for a fixed $\tau$. 
\if\extended 0 
A detailed proof is provided in \cite{extendedITW2022-Gilani}.
\fi
\if\extended 1
A detailed proof is in Section~\ref{subsec:remark1}.
\fi
\end{remark}
%%%%%%

\section{Relationship between maximal $\alpha,\beta$-leakage, and other measures}\label{sec:relationship}
\begin{figure}%[h]
\centering
\includegraphics[width=0.4\textwidth]{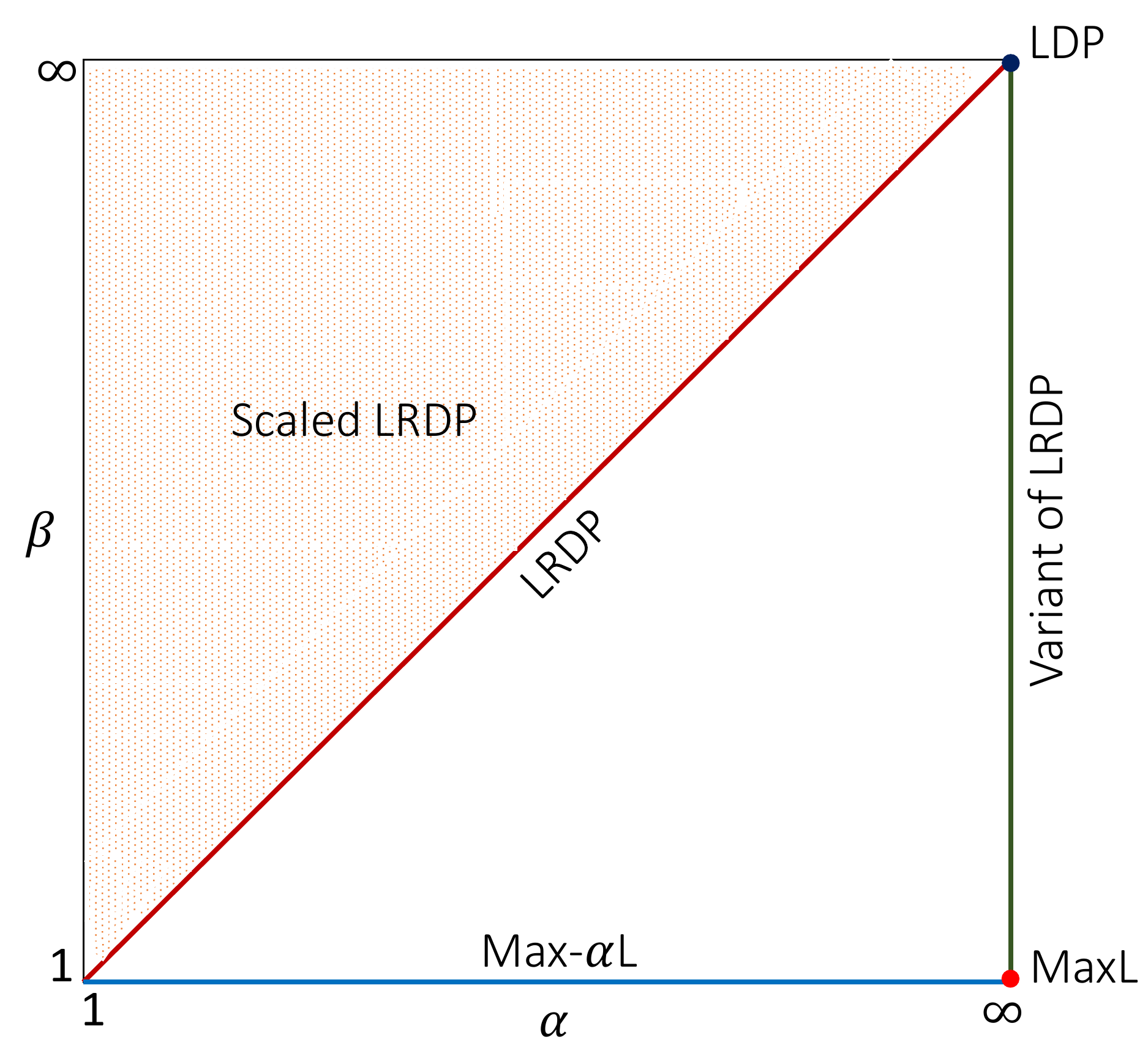}
\centering
\caption{Relationship between maximal $\alpha,\beta$-leakage and other leakage measures as a function of $\alpha$ and $\beta$. For $\alpha  \le\beta$, we obtain a scaled LRDP of order $\beta$. For $\alpha=\beta=\infty$, we obtain LDP. For $\beta=1$, we obtain maximal $\alpha$-leakage which simplifies to maximal leakage for $\alpha=\infty$. Finally, for $\alpha=\infty$ and arbitrary $\beta$, we obtain a variant of LRDP. }%   In particular, the family of measures includes as special cases maximal $\alpha$-leakage (Max-$\alpha$L), maximal leakage (MaxL), local differential privacy (LDP), and local R\'enyi differential privacy (LRDP).}
\label{fig:alpha_beta_leakage_and_other_measures}
\end{figure}
% \section{Relationship between $\alpha,\beta$-Leakage and Other Information Leakage Measures}\label{sec:relationship}
%Maximal $\alpha,\beta$-leakage includes maximal $\alpha$-leakage as a special case for $\beta=1$. Moreover, the choice of $\beta$ helps to recover other leakage measures such as LDP for $\alpha=\infty$ and $\beta=\infty$ and local R\'{enyi} DP for $\alpha=\beta$. The relationships of maximal $\alpha,\beta$-leakage to other measures is discussed in details in section \ref{sec:relationship}.
%%%%
% We now explore the relationship of maximal $\alpha,\beta$-leakage to other measures. 
% Maximal $\alpha,\beta$-leakage includes maximal $\alpha$-leakage as a special case for $\beta=1$.
As mentioned earlier, maximal $\alpha,\beta$-leakage recovers maximal $\alpha$-leakage (and thus maximal leakage) when $\beta=1$. The choices of  $\alpha$ and $\beta$ help to recover other leakage measures such as a scaled LRDP for $\alpha\leq \beta$, LRDP for $\alpha=\beta$, LDP for $(\alpha=\infty, \beta=\infty)$, and a variant of LRDP for $\alpha=\infty$ and arbitrary $\beta$, as shown in Fig.~\ref{fig:alpha_beta_leakage_and_other_measures}. We now present these in detail. %Details are in order.

When $\alpha\leq \beta$, we have
\begin{align}
\nonumber&\mathcal{L}_{\alpha,\beta}(X\to Y)\\ \nonumber 
&=\max_{x'} \  \sup_{P_{\Tilde{X}}}\frac{\alpha}{(\alpha-1)\beta}\log 
\sum_y P_{Y|X}(y|x')^{1-\beta}  \\\label{eq:convexity}&  \hspace{5mm}\left(\sum_{x} P_{\Tilde{X}}(x) P_{Y|X}(y|x)^\alpha \right)^{\beta/\alpha}
\\\label{eq:simplified}
&=\max_{x'} \ \max_x \frac{\alpha}{(\alpha-1)\beta}
 \log \sum_y P_{Y|X}(y|x')^{1-\beta} P_{Y|X}(y|x)^\beta
\end{align}
where \eqref{eq:simplified} follows because the argument of the logarithm in \eqref{eq:convexity} is convex in $P_{\Tilde{X}}$ and so the supremum is attained at an extreme point. This quantity is a scaled LRDP of order $\beta$. Furthermore, if $\alpha=\beta$, the expression in \eqref{eq:simplified} reduces to
\begin{align}\label{beta_beta}
&\nonumber\mathcal{L}_{\beta,\beta}(X\to Y) \\
&=\max_{x'} \ \max_x \frac{1}{\beta-1}
 \log \sum_y P_{Y|X}(y|x')^{1-\beta} P_{Y|X}(y|x)^\beta, 
\end{align}
which is exactly LRDP of order $\beta$. Taking a limit as $\beta\to\infty$ in \eqref{beta_beta} gives %If $\alpha=\infty$ and $\beta=\infty$, then
\begin{align}
\mathcal{L}_{\infty,\infty}(X\to Y)
&=\max_{x'} \, \max_x \, \log\left( \max_y \frac{P_{Y|X}(y|x)}{P_{Y|X}(y|x')}\right)
\\
&=\max_{x,x',y}\, \log \frac{P_{Y|X}(y|x)}{P_{Y|X}(y|x')}
\end{align}
which is LDP. So $\mathcal{L}_{\infty,\beta}(X\to Y)$ passes from maximal leakage at $\beta=1$ to LDP at $\beta=\infty$.
% We start with $\beta=1$, then we have
% \begin{align}
% &\mathcal{L}_{\alpha,1}(X\to Y)\nonumber\\
% &=\sup_{P_{\Tilde{X}}} \frac{\alpha}{\alpha-1} \log \sum_y \left(\sum_x P_{\Tilde{X}}(x) P_{Y|X}(y|x)^\alpha\right)^{1/\alpha}\\
% &=\sup_{P_{\tilde{X}}}I_\alpha^{\text{S}}(\tilde{X};Y),
% \end{align}
% which of course is just maximal $\alpha$-leakage by \eqref{eqn:maximalalpha}. Furthermore, if $\beta=1$ and $\alpha=\infty$, then we have
% \begin{equation}
% \mathcal{L}_{\infty,1}(X\to Y)
% =\log \sum_y \max_x P_{Y|X}(y|x)
% \end{equation}
% which of course is maximal leakage.

From Theorem \ref{thm:alpha-beta-leakage}, for $\alpha=\infty$ and arbitrary $\beta$,  we obtain
\begin{align}
&\nonumber\mathcal{L}_{\infty,\beta}(X\to Y)
\\&=  \max_{x'}  \frac{1}{\beta} \log
\sum_y P_{Y|X}(y|x')^{1-\beta} \max_x P_{Y|X}(y|x)^\beta.
\end{align}
This quantity is a variant of LRDP, and it differs from LRDP of order $\beta$ only in that the max over $x$ is inside the summation over $y$ rather than outside. As far as we know, this quantity has not appeared before.

Finally, we propose an extension of maximal $\alpha,\beta$-leakage so as to include the data of multiple users, shown in \eqref{n_length_extension}. Here, $X^n=(X_1,X_2,\ldots,X_n)$ is a dataset with $n$ entries, $X_{-i}$ represents all entries except the $i$th, and as usual $Y$ is the released data. Thus, the measure in \eqref{n_length_extension} characterizes a situation in which an adversary may have side information access to all entries of the dataset except one. It is clear that this measure collapses to maximal $\alpha,\beta$-leakage for $n=1$. We conjecture that, for $n>1$, it recovers (non-local) DP~\cite{Dwork2006} $(\alpha=\infty,\beta=\infty)$ and RDP~\cite{mironov2017renyi} $(\alpha=\beta)$ for multi-user data across neighbouring databases.

\addtocounter{equation}{1}

\section{Proofs}\label{sec:proofs}
\if \extended 0
Due to space limitations, some proofs have been omitted. Complete proofs can be found in \cite{extendedITW2022-Gilani}.
\fi

\subsection{Proof of Theorem~\ref{thm:alpha-beta-leakage}}\label{proof:thm1}
For $\alpha\in (1,\infty)$ and $\beta \in [1,\infty)$, we first bound $\mathcal{L}_{\alpha,\beta}(X \to Y)$ from above and then, give an achievable scheme.\\
\textbf{Upper Bound:}
Consider the denominator of \eqref{eqn:alpha,beta-leakage-original-def}:
\begin{align}
\max_{P_{\hat{U}}} \sum_u P_U(u)P_{\hat{U}}(u)^{\frac{\alpha-1}{\alpha}}.
\end{align}
This is solved by $P_U(u) P_{\hat{U}}(u)^{-1/\alpha}=\nu$, 
%\begin{equation}
%P_U(u) P_{\hat{U}}(u)^{-1/\alpha}=\nu
%\end{equation}
for some constant $\nu$. So we have
\begin{align}
P_{\hat{U}}(u)=\frac{P_U(u)^\alpha}{\sum_{u'} P_U(u')^\alpha}.
\end{align}
Thus the denominator becomes
\begin{align}
\sum_u P_U(u) \left(\frac{P_U(u)^\alpha}{\sum_{u'} P_U(u')^\alpha}\right)^{\frac{\alpha-1}{\alpha}}=\left(\sum_u P_U(u)^\alpha\right)^{\frac{1}{\alpha}}.
\end{align}
Similarly, the numerator becomes
\begin{align}
\left[\sum_y P_Y(y) \left(\sum_u P_{U|Y}(u|y)^\alpha\right)^{\beta/\alpha}\right]^{1/\beta}.
\end{align}
Thus, the logarithmic term in \eqref{eqn:alpha,beta-leakage-original-def} reduces to
\begin{align}
& \log\frac{\left[\sum_y P_Y(y) \left(\sum_u P_{U|Y}(u|y)^\alpha\right)^{\beta/\alpha}\right]^{1/\beta}}
{\left(\sum_u P_U(u)^\alpha\right)^{1/\alpha}}
\\&= \log
\frac{\left[
\sum_y P_Y(y)^{1-\beta} \left(\sum_u P_{U,Y}(u,y)^\alpha\right)^{\beta/\alpha}
\right]^{1/\beta}}
{\left(\sum_u P_U(u)^\alpha\right)^{1/\alpha}}
\\\label{eq:simplified-leakage}
&=\frac{1}{\beta}\log
\sum_y P_Y(y)^{1-\beta} \left[\frac{\sum_u P_U(u)^\alpha P_{Y|U}(y|u)^\alpha}{\sum_u P_U(u)^\alpha}\right]^{\frac{\beta}{\alpha}}.
\end{align}
Using Jensen's inequality and the Markov chain $U-X-Y$, we have
\begin{align}
P_{Y|U}(y|u)^\alpha &= \left(\sum_{x} P_{X|U}(x|u)P_{Y|X}(y|x)\right)^\alpha\\
&\le \sum_{x} P_{X|U}(x|u) P_{Y|X}(y|x)^\alpha.
\end{align}
%Thus, we may upper bound \eqref{eq:simplified-leakage} by
%\begin{align}
%\nonumber&\frac{1}{\beta}\log
%\sum_y P_Y(y)^{1-\beta} \\&\left[\frac{\displaystyle\sum_{u,x} P_{U}(u)^\alpha P_{X|U}(x|u) P_{Y|X}(y|x)^\alpha}{\displaystyle\sum_u P_U(u)^\alpha} \right]^{\frac{\beta}{\alpha}}.
%\end{align}
So maximal $\alpha,\beta$-leakage may be bounded from above by
\begin{align}
%&\nonumber\mathcal{L}_{\alpha,\beta}(X \to Y)\\\nonumber\leq
&\nonumber\sup_{P_{X}}\sup_{U\to X\to Y} \frac{\alpha}{(\alpha-1)\beta}
\log
\sum_y P_Y(y)^{1-\beta} \\&\hspace{5mm}\left[\frac{\displaystyle\sum_{u,x} P_{U}(u)^\alpha P_{X|U}(x|u) P_{Y|X}(y|x)^\alpha}{\sum_u P_U(u)^\alpha} \right]^{\frac{\beta}{\alpha}}
\\\nonumber
\leq& \sup_{P_{X}}\ \sup_{P_{\Tilde{X}}} \frac{\alpha}{(\alpha-1)\beta}\log
\sum_y P_Y(y)^{1-\beta} \Bigg[\sum_{x} P_{\Tilde{X}}(x)\\&\hspace{5mm} P_{Y|X}(y|x)^\alpha \Bigg]^{\frac{\beta}{\alpha}}\label{eq:upper-bound}
\end{align}
where
\begin{equation}
P_{\Tilde{X}}(x)=\frac{\sum_{u} P_{U}(u)^\alpha P_{X|U}(x|u)}{\sum_u P_U(u)^\alpha}.
\end{equation}
%%%%
\textbf{Lower Bound:} The proof is based on the expression  in \eqref{eq:simplified-leakage} as well as ``shattering'' method. Consider  a random variable $U$ such that $U\to X \to Y$ form a Markov chain and $H(X|U)=0$. For each $x$, let $\mathcal{U}_{x}$ be a finite set such that  $U=u \in \mathcal{U}_{x}$ if and only if $X=x$ and $\mathcal{U}=\bigcup_{x \in \mathcal{X}} \mathcal{U}_{x}$. Moreover, given $X=x$ let $U$ be uniformly distributed on $\mathcal{U}_{x}$.  That is, 
\begin{align}
 P_{U|X}(u|x)=\begin{cases} \displaystyle \frac{1}{|\mathcal{U}_{x}|} & \text{for all}~ u \in \mathcal{U}_{x} \\ 0 & \text{otherwise,}\end{cases}   
\end{align}
and so
\begin{align}
P_{Y|U}(y|u)=\begin{cases} P_{Y|X}(y|x) & \text{for all}~ u \in \mathcal{U}_{x} \\ 0 & \text{otherwise.}\end{cases}
\end{align}
Therefore, we have
\begin{align}
   &\frac{\sum_u P_U(u)^\alpha P_{Y|U}(y|u)^\alpha}{\sum_u P_U(u)^\alpha}\\
   &=\frac{\sum_{x \in \mathcal{X}}\sum_{u\in \mathcal{U}_{x}} \left( \displaystyle\frac{P_X(x) P_{U|X}(u|x)}{P_{X|U}(x|u)}\right)^\alpha P_{Y|U}(y|u)^\alpha}{\sum_{x \in \mathcal{X}}\sum_{u\in \mathcal{U}_{x}}\left( \displaystyle\frac{P_X(x) P_{U|X}(u|x)}{P_{X|U}(x|u)}\right)^\alpha}\\
   &=\frac{\sum_{x} |\mathcal{U}_{x}|^{1-\alpha} P_{X}(x)^\alpha P_{Y|X}(y|x)^\alpha}{\sum_{x} |\mathcal{U}_x|^{1-\alpha} P_{X}(x)^\alpha}.
\end{align}
So we may bound maximal $\alpha,\beta$-leakage from below by
\begin{align}
%\nonumber&\mathcal{L}_{\alpha,\beta}(X\to Y)\\\nonumber&\geq
&\nonumber\sup_{P_{X}} \ \sup_{\mathcal{U}_x}\frac{\alpha}{(\alpha-1)\beta}
\ \log \sum_y P_Y(y)^{1-\beta}\\&\hspace{5mm}\left(\frac{\sum_{x} |\mathcal{U}_{x}|^{1-\alpha} P_{X}(x)^\alpha P_{Y|X}(y|x)^\alpha}{\sum_{x} |\mathcal{U}_x|^{1-\alpha} P_{X}(x)^\alpha}\right)^{\frac{\beta}{\alpha}}\\
\label{lower-bound}&=\sup_{\substack{P_{X},\\P_{\Tilde{X}}}} \frac{\alpha}{(\alpha-1)\beta}
\log
\sum_y P_Y(y)^{1-\beta} \big(\sum_{x} P_{\Tilde{X}}(x) P_{Y|X}(y|x)^\alpha \big)^{\frac{\beta}{\alpha}}
\end{align}
where here $P_{\Tilde{X}}(x)=\displaystyle\frac{|\mathcal{U}_{x}|^{1-\alpha}P_{X}(x)^\alpha}{\sum_{x} |\mathcal{U}_x|^{1-\alpha} P_{X}(x)^\alpha},$
%\begin{equation}
%P_{\Tilde{X}}(x)=\frac{|\mathcal{U}_{x}|^{1-\alpha}P_{X}(x)^\alpha}{\sum_{x} |\mathcal{U}|^{1-\alpha} P_{X}(x)^\alpha},
%\end{equation}
and we have used the fact that any distribution $P_{\Tilde{X}}(x)$ can be reached with appropriate choice of $|\mathcal{U}_{x}|$, assuming $P_{X}(x)>0$ for all $x$; this condition can be assumed because any $P_{X}$ is arbitrarily close to a distribution with full support. Thus, combining \eqref{eq:upper-bound} and \eqref{lower-bound}, we have
\begin{align}\label{result thm1}
&\nonumber\mathcal{L}_{\alpha,\beta}(X\to Y)=\sup_{P_X}\ \sup_{P_{\Tilde{X}}} \frac{\alpha}{(\alpha-1)\beta}
\\
&\log
\sum_y P_Y(y)^{1-\beta} \left(\sum_{x} P_{\Tilde{X}}(x) P_{Y|X}(y|x)^\alpha \right)^{\frac{\beta}{\alpha}}.
\end{align}
Since the choice of $P_X$ only impacts $P_Y$, and the supremum of a convex function is attained at an extreme point, we may simplify \eqref{result thm1} as follows.
\begin{align*}
&\nonumber\mathcal{L}_{\alpha,\beta}(X\to Y)
=\max_{x'} \  \sup_{P_{\Tilde{X}}}\frac{\alpha}{(\alpha-1)\beta} \\&   \log \sum_y P_{Y|X}(y|x')^{1-\beta} \left(\sum_{x} P_{\Tilde{X}}(x) P_{Y|X}(y|x)^\alpha \right)^{\beta/\alpha}.
\end{align*}
%%%
\subsection{Proof of Theorem~\ref{thm2}}\label{proof:thm2}
\textbf{Monotonicity in $\beta$:} For $\alpha\in (1,\infty)$, $\beta_1,\beta_2 \in [1,\infty)$ and $\beta_2 > \beta_1$, consider the argument of the logarithm in \eqref{thm1}:
\begin{align}
& 
\sum_y P_{Y|X}(y|x')^{1-\beta_1}\left[\sum_{x} P_{\Tilde{X}}(x) P_{Y|X}(y|x)^\alpha\right]^{\frac{\beta_1}{\alpha}}\\
\nonumber &=
 \sum_{y}P_{Y|X}(y|x')\Bigg[ P_{Y|X}(y|x')^{-\alpha}\sum_{x} P_{\Tilde{X}}(x
    ) \\&\hspace{10mm}P_{Y|X}(y|x)^{\alpha}\Bigg]^{\frac{\beta_2\beta_1}{\alpha\beta_2}}\\\nonumber
 &\leq \Bigg[
\sum_{y} P_{Y|X}(y|x')\bigg(P_{Y|X}(y|x')^{-\alpha}\sum_{x} P_{\Tilde{X}}(x
    ) \\&\hspace{10mm}P_{Y|X}(y|x)^{\alpha}\bigg)^{\frac{\beta_2}{\alpha}}\Bigg]^{ \frac{\beta_1}{\beta_2}}\\
 &= \Bigg[
\sum_{y} P_{Y|X}(y|x')^{1-\beta_2}\bigg(\sum_{x} P_{\Tilde{X}}(x
    ) P_{Y|X}(y|x)^{\alpha}\bigg)^{\frac{\beta_2}{\alpha}}\Bigg]^{ \frac{\beta_1}{\beta_2}}
    \end{align}
    where the inequality results from applying Jensen's inequality to the concave function $f:\;x\to x^{p}\;(x \geq 0,\; p< 1)$. For $\alpha \in (1,\infty)$ and $\beta \in [1,\infty)$, the function $f: t \to \frac{\alpha}{(\alpha-1)\beta} \log t$ is increasing in $t>0$. Therefore, we have
    \begin{align}
&\nonumber \frac{\alpha}{(\alpha-1)\beta_1}  \log \sum_y P_{Y|X}(y|x')^{1-\beta_1} \bigg[\sum_{x} P_{\Tilde{X}}(x) \\&\hspace{10mm} P_{Y|X}(y|x)^\alpha \bigg]^{\frac{\beta_1}{\alpha}}
\\\nonumber\leq & \frac{\alpha}{(\alpha-1)\beta_2}\log\sum_{y} P_{Y|X}(y|x')^{1-\beta_2}\bigg[ \sum_{x} P_{\Tilde{X}}(x
    ) \\&\hspace{10mm}P_{Y|X}(y|x)^{\alpha}\bigg]^{\frac{\beta_2}{\alpha}}.
\end{align}
Taking the max over $x^\prime$ and supremum over $P_{\Tilde{X}}$ completes the proof. Another way to prove this property is to consider the numerator in \eqref{eqn:alpha,beta-leakage-original-def} as the $\beta$-norm of a random variable. Since the $\beta$-norm of a random variable is non-decreasing in $\beta$, maximal $\alpha,\beta$-leakage is non-decreasing in $\beta$.
%%%
\\\textbf{Data processing inequalities:} Let random variables $X,Y,Z$ form a Markov chain, i.e., $X-Y-Z$.
\if \extended 0
First consider the post-processing inequality, that is,
\begin{equation}
        \mathcal{L}_{\alpha,\beta}(X\to Z) \leq \mathcal{L}_{\alpha,\beta}(X \to Y).
\end{equation}
The proof of this, contained in the extended version \cite{extendedITW2022-Gilani}, is based on the expression for maximal $\alpha,\beta$-leakage in \eqref{thm1}, and proper use of Jensen's inequality.
\fi
\if \extended 1
Based on the expression of maximal $\alpha,\beta$-leakage in \eqref{result thm1} we first prove that 
\begin{align}
        \mathcal{L}_{\alpha,\beta}(X\to Z) \leq \mathcal{L}_{\alpha,\beta}(X \to Y).
\end{align}
 For any $y \in \mathcal{Y}$, let
\begin{align}\label{def:g}
    g(y)=\left( \sum_{x} P_{\Tilde{X}}(x) P_{Y|X}(y|x)^{\alpha}\right)^{\frac{1}{\alpha}}
\end{align}
and
\begin{align}\label{def:c}
    c_z(y)=\displaystyle\frac{P_{Y}(y)\; P_{Z|Y}(z|y)}{P_{Z}(z)}
\end{align}
such that $\sum_y c_z(y)=1$. We have
\begin{align}
    &\sum_y P_Y(y)^{1-\beta} \left( \sum_{x} P_{\Tilde{X}}(x) P_{Y|X}(y|x)^{\alpha}\right)^{\frac{\beta}{\alpha}}
    \\&=\sum_y P_{Y}(y)^{1-\beta}g(y)^\beta
    \\&=\sum_{y,z} P_Y(y) P_{Z|Y}(z|y)\left(\frac{g(y)}{P_Y(y)}\right)^\beta
    \\&=\sum_{z} P_Z(z) \sum_y c_z(y)\left(\frac{g(y)}{P_Y(y)}\right)^\beta 
    \\&\ge \sum_z P_Z(z) \left(\sum_y c_z(y)\frac{g(y)}{P_Y(y)}\right)^\beta \label{jensens_application}
    \\&=\sum_z P_{Z}(z)^{1-\beta} \left(\sum_y P_{Z|Y}(z|y)g(y)\right)^\beta\label{g_form}
\end{align}
where \eqref{jensens_application} follows from applying Jensen's inequality to the convex function $f:\;x\to x^{p}\;(x \geq 0,\; p\geq 1)$.
% , we have
% \begin{align}\label{ineq:jensen}
% \displaystyle \sum_y c_z(y)\left(\frac{g(y)}{P_{Y}(y)}\right)^\beta \geq
% \left(\displaystyle \sum_y c_z(y)\frac{g(y)}{P_{Y}(y)}\right)^\beta, 
% \end{align}
% and using the definition of $c_z(y)$ in \eqref{def:c} we get
% %
% \begin{align}
% &\nonumber\displaystyle \sum_y P_{Y}(y)\; P_{Z|Y}(z|y)\left(\frac{g(y)}{P_{Y}(y)}\right)^\beta
% \\&\geq
%  P_{Z}(z) \left(\displaystyle \sum_y \frac{P_{Z|Y}(z|y)}{P_{Z}(z)} g(y)\right)^\beta. 
% \end{align}
% We now take a summation over $z$. Thus,
% \begin{align}
% &\nonumber \displaystyle \sum_z \displaystyle \sum_{y}P_{Y}(y)\; P_{Z|Y}(z|y)\left(\frac{g(y)}{P_{Y}(y)}\right)^\beta 
% \\ &\geq
% \displaystyle \sum_z P_{Z}(z) \left(\displaystyle \sum_y \frac{P_{Z|Y}(z|y)}{P_{Z}(z)} g(y)\right)^\beta.
% \end{align}
% Therefore, 
% \begin{align}
%  &\nonumber\displaystyle \sum_y P_{Y}(y)^{1-\beta}g(y)^\beta
% \\ &\geq
%  \displaystyle\sum_z P_{Z}(z)^{1-\beta} \left(\sum_y P_{Z|Y}(z|y)g(y)\right)^\beta. 
% \end{align}
Recalling the definition of $g(y)$ from \eqref{def:g}, we have
\begin{align}
    &\sum_y P_{Z|Y}(z|y) g(y)
    \\&=\sum_y P_{Z|Y}(z|y) \bigg( \sum_{x} P_{\Tilde{X}}(x)  P_{Y|X}(y|x)^{\alpha}\bigg)^{\frac{1}{\alpha}}
    \\&=\sum_y \bigg(\sum_{x}  \Big( P_{\Tilde{X}}(x)^{\frac{1}{\alpha}}P_{Z|Y}(z|y)  P_{Y|X}(y|x)\Big)^{\alpha}\bigg)^{\frac{1}{\alpha}}
    \\&\ge \bigg(\sum_{x}  \Big( \sum_y  P_{\Tilde{X}}(x)^{\frac{1}{\alpha}}P_{Z|Y}(z|y)  P_{Y|X}(y|x)\Big)^{\alpha}\bigg)^{\frac{1}{\alpha}}\label{ineq:norm}
    \\&=\left(\sum_{x} P_{\Tilde{X}}(x)  P_{Z|X}(z|x)^{\alpha}\right)^{\frac{1}{\alpha}}\label{applying Markov chain}
\end{align}
% Thus, using the definition of $g(y)$ in \eqref{def:g} we have
% %
% \begin{align}\label{ineq:plugging in g}
%  &\nonumber\displaystyle \sum_y P_{Y}(y)^{1-\beta}\left( \sum_{x} P_{\Tilde{X}}(x) P_{Y|X}(y|x)^{\alpha}\right)^{\frac{\beta}{\alpha}}
% \\\nonumber \geq &
%  \displaystyle\sum_z P_{Z}(z)^{1-\beta} \Bigg(\sum_y P_{Z|Y}(z|y) \bigg( \sum_{x} P_{\Tilde{X}}(x) \\&\hspace{10mm} P_{Y|X}(y|x)^{\alpha}\bigg)^{\frac{1}{\alpha}}\Bigg)^\beta 
%  %
%  \\\nonumber
%  =&\displaystyle \sum_z P_{Z}(z)^{1-\beta} \Bigg(\displaystyle\sum_y \bigg(\displaystyle\sum_{x}  \Big( P_{\Tilde{X}}(x)^{\frac{1}{\alpha}}P_{Z|Y}(z|y) \\&\hspace{10mm} P_{Y|X}(y|x)\Big)^{\alpha}\bigg)^{\frac{1}{\alpha}}\Bigg)^\beta 
%  %
%  \\\nonumber\label{ineq:norm}\geq & 
% \displaystyle \sum_z P_{Z}(z)^{1-\beta} \Bigg(\bigg(\displaystyle\sum_{x}  \Big( \displaystyle\sum_y  P_{\Tilde{X}}(x)^{\frac{1}{\alpha}}P_{Z|Y}(z|y) \\&\hspace{10mm} P_{Y|X}(y|x)\Big)^{\alpha}\bigg)^{\frac{1}{\alpha}}\Bigg)^\beta
% %
% \\\nonumber
% = &
% \displaystyle \sum_z P_{Z}(z)^{1-\beta} \bigg(\displaystyle\sum_{x} P_{\Tilde{X}}(x) \Big( \displaystyle\sum_y P_{Z|Y}(z|y) \\&\hspace{10mm} P_{Y|X}(y|x)\Big)^{\alpha}\bigg)^{\frac{\beta}{\alpha}}
% %
% \\\label{applying Markov chain} = & 
% \displaystyle \sum_z P_{Z}(z)^{1-\beta} \left(\sum_{x} P_{\Tilde{X}}(x)  P_{Z|X}(z|x)^{\alpha}\right)^{\frac{\beta}{\alpha}}
% \end{align}
where 
\begin{itemize}
    \item  \eqref{ineq:norm} follows because $p$-norm satisfies the triangle inequality for $p\in (1,\infty)$,
    \item \eqref{applying Markov chain} follows because the Markov chain $X - Y - Z$ holds.
\end{itemize}
Applying \eqref{applying Markov chain} to \eqref{g_form}, and using the fact that for $\alpha \in (1,\infty)$ and $\beta \in [1,\infty)$, the function $f: t \to \frac{\alpha}{(\alpha-1)\beta} \log t$ is increasing in $t > 0$, gives
\begin{align}
   & \frac{\alpha}{(\alpha-1)\beta} \log \sum_y P_{Y}(y)^{1-\beta}\left( \sum_{x} P_{\Tilde{X}}(x) P_{Y|X}(y|x)^{\alpha}\right)^{\frac{\beta}{\alpha}}\nonumber\\ \geq & \frac{\alpha}{(\alpha-1)\beta} \log  \sum_z P_{Z}(z)^{1-\beta} \left(\sum_{x} P_{\Tilde{X}}(x)  P_{Z|X}(z|x)^{\alpha}\right)^{\frac{\beta}{\alpha}}.
\end{align}
Taking suprema over $P_X$ and $P_{\Tilde{X}}$ completes the proof.
\fi

We now prove the linkage inequality, that is
\begin{align}
      \mathcal{L}_{\alpha,\beta}(X\to Z) \leq \mathcal{L}_{\alpha,\beta}(Y \to Z),
\end{align}
using the definition of maximal $\alpha,\beta$-leakage in \eqref{eqn:alpha,beta-leakage-original-def}. 
\if \extended 0
Let $f(P_{UY})$ be the quantity inside the suprema in \eqref{eqn:alpha,beta-leakage-original-def}, which depends only on the joint distribution of $U$ and $Y$. Similarly, we write $f(P_{UZ})$ for the same quantity when $Y$ is replaced by $Z$.
\fi
\if \extended 1
Let
\begin{align}
    \nonumber &f(P_{UZ})=\frac{\alpha}{\alpha-1} \log \\& \frac{\displaystyle \max_{P_{\hat{U}|Z}} \left[\sum_z P_Z(z) \left(\sum_u P_{U|Z}(u|z) P_{\hat{U}|Z}(u|z)^{\frac{\alpha-1}{\alpha}}\right)^{\beta}\right]^{1/\beta}}{\displaystyle \max_{P_{\hat{U}}} \sum_u P_U(u)P_{\hat{U}}(u)^{\frac{\alpha-1}{\alpha}}}.
\end{align}
\fi
For the Markov chain $X-Y-Z$, we have
\begin{align}
  \mathcal{L}_{\alpha,\beta}(X\to Z)&=\sup_{P_{X}} \ \sup_{U\to X\to Z}  \ f(P_{UZ})
\\\label{expand_Markov}
 &= \sup_{P_{X}} \ \sup_{U\to X \to Y\to Z } f(P_{UZ})
\\
 &\le  \sup_{P_{X}} \ \sup_{U\to Y \to Z } f(P_{UZ})\\\label{P_x to P_y}
 &\le  \sup_{P_{Y}} \ \sup_{U\to Y\to Z} f(P_{UZ})\\\nonumber
 &=\mathcal{L}_{\alpha,\beta}(Y\to Z)
\end{align}
where \eqref{expand_Markov} follows because $P_{UZ}$ are the same under the Markov chains $U-X-Z$ and $U-X-Y-Z$, and \eqref{P_x to P_y} follows from the fact that a subset of all distributions $P_{Y}$ is reachable from the distribution $P_{X}$.
%%%
\\
\textbf{Non-negativity:} Consider the logarithmic term in \eqref{thm1}:
\begin{align}
 & \log 
\sum_y P_{Y|X}(y|x')^{1-\beta} \left(\sum_{x} P_{\Tilde{X}}(x) P_{Y|X}(y|x)^\alpha \right)^{\frac{\beta}{\alpha}}\\
\label{jensen1}
&\geq  \log
\sum_{y} P_{Y|X}(y|x')^{1-\beta}\bigg(\sum_{x}  P_{\Tilde{X}}(x) P_{Y|X}(y|x)\bigg)^{\beta}\\
&=\log
\sum_{y} P_{Y|X}(y|x')\left(\frac{\sum_{x} P_{\Tilde{X}}(x) P_{Y|X}(y|x)}{P_{Y|X}(y|x')}\right)^{\beta}\\
\label{jensen2}
&\geq \log
 \left(\sum_{y} P_{Y|X}(y|x') \ \frac{\sum_{x} P_{\Tilde{X}}(x) P_{Y|X}(y|x)}{P_{Y|X}(y|x')}\right)^{\beta}\\
&=\log
\bigg(\sum_{x,y} P_{\Tilde{X}}(x) P_{Y|X}(y|x)\bigg)^{\beta}=\log 1 =0
\end{align}
where both inequalities follow from applying Jensen's inequality to the convex function $f:\;x\to x^{p}\;(x \geq 0,\; p\geq1)$ and the fact that logarithmic functions are increasing. Equality holds in the first inequality if and only if for any $y \in \mathcal{Y}$, $P_{Y|X}(y|x)$ are the same for all $x\in \mathcal{X}$. Thus, we have
\begin{align}
    P_{Y|X}(y|x)=P_{Y}(y) \quad x\in \mathcal{X}, y \in \mathcal{Y}
\end{align}
which means $X$ and $Y$ are independent. This condition is also sufficient for equality in the second inequality. 
%%%%%%%
\\
\textbf{Additivity:}
\if \extended 0
Due to space limitations, the proof is relegated to \cite{extendedITW2022-Gilani}.
\fi
\if \extended 1
We first prove additivity for $n=2$. We have $P_{X_1Y_1X_2Y_2}=P_{X_1Y_1}\cdot P_{X_2Y_2}$. To prove the additivity in \eqref{eqn:additivity}, using Theorem~\ref{thm:alpha-beta-leakage} it suffices to show that
\begin{align}\label{eqn:additivityproof1}
   &\sup_{P_{\tilde{X}_1,\tilde{X}_2}} \sum_{y_1,y_2} P_{Y_1Y_2|X_1X_2}(y_1,y_2|x_1',x_2')^{1-\beta}\nonumber\\
   &\left(\sum_{x_1,x_2} P_{\Tilde{X}_1,\Tilde{X}_2}(x_1,x_2) P_{Y_1Y_2|X_1X_2}(y_1,y_2|x_1,x_2)^\alpha \right)^{\beta/\alpha}\nonumber\\
   &=\sup_{\substack{P_{\tilde{X}_i}\\ i\in{1,2}}}\prod_{i=1}^2\Big(\sum_{y_i} P_{Y|X}(y_i|x_i')^{1-\beta} \big(\sum_{x_i} P_{\Tilde{X_i}}(x_i)\nonumber\\
   &\hspace{1cm}P_{Y_i|X_i}(y_i|x_i)^\alpha \big)^{\beta/\alpha}\Big),
\end{align}
for every $x_1',x_2'$.
We simplify LHS in \eqref{eqn:additivityproof1} as 
\begin{align}
    &\sup_{P_{\tilde{X}_1,\tilde{X}_2}} \sum_{y_1,y_2} P_{Y_1Y_2|X_1X_2}(y_1,y_2|x_1',x_2')^{1-\beta}\nonumber\\
   &\Big(\sum_{x_1,x_2} P_{\Tilde{X}_1,\Tilde{X}_2}(x_1,x_2) P_{Y_1Y_2|X_1X_2}(y_1,y_2|x_1,x_2)^\alpha \Big)^{\beta/\alpha}\nonumber\\
   =&\sup_{P_{\tilde{X}_1,\tilde{X}_2}} \sum_{y_1,y_2} P_{Y_1|X_1}(y_1|x_1')^{1-\beta}P_{Y_2|X_2}(y_2|x_2')^{1-\beta}\nonumber\\
   &\Big(\sum_{x_1,x_2} P_{\Tilde{X}_1,\Tilde{X}_2}(x_1,x_2) P_{Y_1|X_1}(y_1|x_1)^\alpha P_{Y_2|X_2}(y_2|x_2)^\alpha \Big)^{\beta/\alpha}\label{eqn:additivityproof2}.
\end{align}
Let $k(y_1)=\sum_{x_1}P_{\tilde{X}_1}(x_1)P_{Y_1|X_1}(y_1|x_1)^\alpha$, for all $y_1$, so that we can define a set of probability distributions over $\mathcal{X}_1$ as
\begin{align}
    P_{\hat{X}_1}(x_1|y_1)=\frac{P_{\tilde{X}_1}(x_1)P_{Y_1|X_1}(y_1|x_1)^\alpha}{k(y_1)}. 
\end{align}
Thus, \eqref{eqn:additivityproof2} is equal to
\begin{align}
    &\sup_{P_{\tilde{X}_1,\tilde{X}_2}} \sum_{y_1,y_2} P_{Y_1|X_1}(y_1|x_1')^{1-\beta}P_{Y_2|X_2}(y_2|x_2')^{1-\beta}\nonumber\\
   &\hspace{1cm}\Big(\sum_{x_1,x_2} k(y_1)P_{\hat{X}_1|Y_1}(x_1|y_1)P_{\tilde{X}_2|\tilde{X}_1}(x_2|x_1)\nonumber\\
   &\hspace{2cm}P_{Y_2|X_2}(y_2|x_2)^\alpha \Big)^{\beta/\alpha}\\
   &\leq\sup_{P_{\tilde{X}_1},P_{\tilde{X}_2|X_1}}\sum_{y_1}P_{Y_1|X_1}(y_1|x_1)^{1-\beta}\big(\sum_{x_1}P_{\tilde{X}_1}(x_1)\nonumber\\
   & P_{Y_1|X_1}(y_1|x_1)^\alpha\big)^{\frac{\beta}{\alpha}}\max_{\tilde{y}_1}\sum_{y_2}P_{Y_2|X_2}(y_2|x_2')^{1-\beta}\nonumber\\
   &\Big(\sum_{x_1,x_2}P_{\hat{X}_1|Y_1}(x_1|\tilde{y}_1)P_{\tilde{X}_2|\tilde{X}_1}(x_2|x_1)P_{Y_2|X_2}(y_2|x_2)^\alpha \Big)^{\beta/\alpha}\\
   &=\sup_{P_{\tilde{X}_1},P_{\tilde{X}_2|X_1}}\sum_{y_1}P_{Y_1|X_1}(y_1|x_1)^{1-\beta}\big(\sum_{x_1}P_{\tilde{X}_1}(x_1)\nonumber\\
   & P_{Y_1|X_1}(y_1|x_1)^\alpha\big)^{\frac{\beta}{\alpha}}\sum_{y_2}P_{Y_2|X_2}(y_2|x_2')^{1-\beta}\nonumber\\
   &\Big(\sum_{x_1,x_2}P_{\hat{X}_1|Y_1}(x_1|{y}_1^*)P_{\tilde{X}_2|\tilde{X}_1}(x_2|x_1)P_{Y_2|X_2}(y_2|x_2)^\alpha \Big)^{\beta/\alpha}\label{eqn:additivityproof3}
\end{align}
We now define $$P_{\hat{X}_2}(x_2)=\sum_{x_1}P_{\hat{X}_1|Y_1}(x_1|y_1^*)P_{X_2|X_1}(x_2|x_1),$$ which is a probability distribution over $\mathcal{X}_2$. Then, \eqref{eqn:additivityproof3} is equal to
\begin{align}
    &\sup_{\substack{P_{\tilde{X}_1},\\ P_{\hat{X}_2}}}\sum_{y_1}P_{Y_1|X_1}(y_1|x_1')^{1-\beta}\big(\sum_{x_1}P_{\tilde{X}_1}(x_1)P_{Y_1|X_1}(y_1|x_1)^\alpha\big)^{\frac{\beta}{\alpha}}\nonumber\\
    &\sum_{y_2}P_{Y_2|X_2}(y_2|x_2')^{1-\beta}\big(\sum_{x_2}P_{\hat{X}_2}(x_2)P_{Y_2|X_2}(y_2|x_2)^\alpha\big)^{\frac{\beta}{\alpha}}\\
    &=\sup_{\substack{P_{\tilde{X}_i}\\ i\in{1,2}}}\prod_{i=1}^2\Big(\sum_{y_i} P_{Y|X}(y_i|x_i')^{1-\beta} \big(\sum_{x_i} P_{\Tilde{X_i}}(x_i)\nonumber\\
   &\hspace{1cm}P_{Y_i|X_i}(y_i|x_i)^\alpha \big)^{\beta/\alpha}\Big).
\end{align}
This proves \eqref{eqn:additivityproof1} as the lower bound part of \eqref{eqn:additivityproof1} is trivial. Thus we have 
\begin{align}\label{n=2case}
    \mathcal{L}_{\alpha,\beta}(X_1,X_2\!\rightarrow\! Y_1,Y_2)\!=\!\mathcal{L}_{\alpha,\beta}(X_1\rightarrow Y_1)\!+\!\mathcal{L}_{\alpha,\beta}(X_2\rightarrow Y_2).
\end{align}
Using \eqref{n=2case} twice, we have 
\begin{align}
    &\mathcal{L}_{\alpha,\beta}(X^3\rightarrow Y^3)\nonumber\\
    &=\mathcal{L}_{\alpha,\beta}(X^2\rightarrow Y^2)+\mathcal{L}_{\alpha,\beta}(X_3\rightarrow Y_3)\\
    &=\mathcal{L}_{\alpha,\beta}(X_1\rightarrow Y_1)+\mathcal{L}_{\alpha,\beta}(X_2\rightarrow Y_2)+\mathcal{L}_{\alpha,\beta}(X_3\rightarrow Y_3).
\end{align}
Similarly, by repeated application of \eqref{n=2case} $(n-1)$ times, we get
\eqref{eqn:additivity}.

%%%%%%%%%%
\fi
\if \extended 1
\subsection{Proof for Remark \ref{remark:reparameterization}}\label{subsec:remark1}
Let $\tau \in [0,1]$ and $\beta=\frac{\alpha}{1-\tau(1-\alpha)}$. We may re-write the expression of maximal $\alpha,\beta$-leakage in \eqref{thm1} in terms of $\alpha$ and $\tau$, as follows.
\begin{align}\label{alpha_tau_version}
\nonumber&\mathcal{L}_{\alpha,\tau}(X\to Y)
=\max_{x'}\ \sup_{P_{\Tilde{X}}} \; \frac{1-\tau(1-\alpha)}{\alpha-1}  \log \sum_y \\& P_{Y|X}(y|x')^{\frac{(1-\tau)(1-\alpha)}{1-\tau(1-\alpha)}}
\left(\sum_x P_{\Tilde{X}}(x)P_{Y|X}(y|x)^\alpha\right)^{\frac{1}{1-\tau(1-\alpha)}}.
\end{align}
We claim that this leakage measure is non-increasing in $\tau$ for a fixed $\alpha$, and non-decreasing in $\alpha$ for a fixed $\tau$. Since $\beta$ is decreasing in $\tau$, the first claim, that the measure is non-increasing in $\tau$, is equivalent to it being non-decreasing in $\beta$, which we have already proved in Section \ref{proof:thm2}. To prove the second claim, we first prove the following lemma, which provides a still other representation of the leakage measure. 
\begin{lemma}\label{alpha_tau}
The measure defined in \eqref{alpha_tau_version} can be represented by
\begin{align}\label{alpha_tau_variational}
\nonumber&\mathcal{L}_{\alpha,\tau}(X\to Y)
=\max_{x'}\,\sup_{P_{\Tilde{X}}}\,\inf_{Q_Y} \frac{1}{\alpha-1} \log \sum_{x,y}\\& P_{\Tilde{X}}(x) P_{Y|X}(y|x)^\alpha \left(Q_Y(y)^\tau P_{Y|X}(y|x')^{1-\tau}\right)^{1-\alpha}.
\end{align}
\end{lemma}
\begin{remark}
Some of the relationships to other measures become clear from this lemma. Namely, if $\tau=1$, then we see the definition of Sibson mutual information as
\begin{align}
\inf_{Q_Y} D_\alpha(P_{XY}\|P_X\times Q_Y).
\end{align}
If $\tau=0$, then we see the definition of LRDP as
\begin{align}
\max_{x,x'} D_\alpha(P_{Y|X=x}\|P_{Y|X=x'}).
\end{align}
\end{remark}
\emph{Proof of Lemma \ref{alpha_tau}}:
Consider any $\gamma\in(-\inf,0]\cup [1,\inf)$, and any constants $C(y)$ for $y\in\mathcal{Y}$. Furthermore, consider the optimization problem
\begin{equation}\label{general_QY_opt}
\inf_{Q_Y} \sum_y C(y) Q_Y(y)^\gamma.
\end{equation}
$\gamma$ is in the range where \eqref{general_QY_opt} is convex in $Q_Y$, so it is solved by setting the derivative of $Q_Y(y)$ to a constant:
\begin{align}
\nu=\frac{\partial}{\partial Q_Y(y)} \sum_y C(y) Q_Y(y)^\gamma=C(y)\gamma Q_Y(y)^{\gamma-1}.
\end{align}
We can see that the optimal choice is therefore
\begin{align}
Q_Y(y)=\frac{C(y)^{1/(1-\gamma)}}{\sum_{y'} C(y')^{1/(1-\gamma)}}.
\end{align}
Thus \eqref{general_QY_opt} becomes
\begin{align}\label{optimized_expression}
\frac{\sum_y C(y) C(y)^{\gamma/(1-\gamma)}}{\left(\sum_{y'} C(y')^{1/(1-\gamma)}\right)^\gamma}
=\left(\sum_y C(y)^{1/(1-\gamma)}\right)^{1-\gamma}.
\end{align}
In our case, we have $\gamma=\tau(1-\alpha)<0$, and
\begin{align}
C(y)=\sum_x P_{\Tilde{X}}(x) P_{Y|X}(y|x)^\alpha P_{Y|X}(y|x')^{(1-\tau)(1-\alpha)}.
\end{align}
Applying the result in \eqref{optimized_expression} to our case, we find that \eqref{alpha_tau_variational} is equal to
\begin{align}
\nonumber&\max_{x'}\,\sup_{P_{\Tilde{X}}}\frac{1}{\alpha-1}\log \bigg[\sum_y \bigg(\sum_x P_{\Tilde{X}}(x) P_{Y|X}(y|x)^\alpha\\&\hspace{10mm} P_{Y|X}(y|x')^{(1-\tau)(1-\alpha)}\bigg)^{\frac{1}{1-\tau(1-\alpha)}}\bigg]^{1-\tau(1-\alpha)}
\\\nonumber&=\max_{x'}\,\sup_{P_{\Tilde{X}}}\frac{1-\tau(1-\alpha)}{\alpha-1}
\log \sum_y P_{Y|X}(y|x')^{\frac{(1-\tau)(1-\alpha)}{1-\tau(1-\alpha)}}\\& \hspace{10mm}\left(\sum_x  P_{\Tilde{X}}(x)P_{Y|X}(y|x)^\alpha\right)^{\frac{1}{1-\tau(1-\alpha)}}
\end{align}
which is precisely \eqref{alpha_tau_version}.
Given Lemma~\ref{alpha_tau}, we prove that $\mathcal{L}_{\alpha,\tau}(X\to Y)$ is non-decreasing in $\alpha$ for a fixed $\tau$ as follows. We may write the objective function in \eqref{alpha_tau_variational} as
\begin{align}
\nonumber&\frac{1}{\alpha-1} \log \sum_{x,y} P_{\Tilde{X}}(x) P_{Y|X}(y|x)\\&\hspace{5mm} \left(\frac{P_{Y|X}(y|x)}{Q_Y(y)^\tau P_{Y|X}(y|x')^{1-\tau}}\right)^{\alpha-1}.
\end{align}
This expression is non-decreasing in $\alpha$ due to the fact that, for any distribution $P_Z$, and any constants $C(z)$,
\begin{align}
\frac{1}{\alpha-1} \log \sum_z P_Z(z) C(z)^{\alpha-1}
\end{align}
is non-decreasing in $\alpha$ for $\alpha>1$.
\fi
\bibliographystyle{IEEEtran}
\if \extended 0
\newpage
\fi
\bibliography{Bibliography}
\end{document}